# THE COLOR AND BINARITY OF (486958) 2014 MU$_{69}$ AND OTHER LONG-RANGE NEW HORIZONS KUIPER BELT TARGETS

S.D. Benecchi[1], D. Borncamp[2], A. Parker[3], M.W. Buie[3], K.S. Noll[4], R.P. Binzel[5], S.A. Stern[3], A.J. Verbiscer[6], J. J. Kavelaars[7,8], A.M. Zangari[3], J. R. Spencer[3], and H.A. Weaver[9]

[1] Planetary Science Institute, 1700 East Fort Lowell, Suite 106, Tucson, AZ 85719; susank@psi.edu
[2] Space Telescope Science Institute, 3700 San Martin Drive, Baltimore, MD 2121; now at Decipher Technology Studios 110 S. Union Street, Floor 2 Alexandria, VA 22314
[3] Southwest Research Institute, 1050 Walnut St., Suite 300, Boulder, CO 80302
[4] NASA Goddard Space Fight Center, 8800 Greenbelt Rd. Code 693, Greenbelt, MD 20771
[5] Massachusetts Institute of Technology, Cambridge MA
[6] Department of Physics and Astronomy, University of Victoria, Elliott Building, 3800 Finnerty Rd, Victoria, BC V8P 5C2, Canada
[7] NRC-Herzberg Astronomy and Astrophysics, National Research Council of Canada, 5071 West Saanich Rd, Victoria, BC V9E 2E7, Canada
[8] University of Virginia, Department of Astronomy, PO Box 400325, Charlottesville, VA 22904, verbiscer@virginia.edu
[9] Space Department, Johns Hopkins University Applied Physics Laboratory, 11100 Johns Hopkins Road, Laurel, MD 20723

Pages: 15
Figures: 5
Tables: 3

**Proposed running head:** The Color of (486958) 2014 MU$_{69}$

**Corresponding author:** S. D. Benecchi, susank@psi.edu, Planetary Science Institute, 1700 East Fort Lowell, Suite 106, Tucson, AZ 85719.

**ABSTRACT**

The Hubble Space Telescope (HST) measured the colors of eight Kuiper Belt Objects (KBOs) that will be observed by the New Horizons spacecraft including its 2019 close fly-by target the Cold Classical KBO (486958) 2014 MU$_{69}$. We find that the photometric colors of all eight objects are red, typical of the Cold Classical dynamical population within which most reside. Because 2014 MU$_{69}$ has a similar color to that of other KBOs in the Cold Classical region of the Kuiper Belt, it may be possible to use the upcoming high-resolution New Horizons observations of 2014 MU$_{69}$ to draw conclusions about the greater Cold Classical population. Additionally, HST found none of these KBOs to be binary within separations of ~0.06 arcsec (~2000 km at 44 AU range) and $\Delta m \leq 0.5$. This conclusion is consistent with the lower fraction of binaries found at relatively wide separations. A few objects appear to have significant photometric variability, but our observations are not of sufficient signal-to-noise or time duration for further interpretation.

*Subject headings*: Kuiper Belt; Photometry; Hubble Space Telescope observations

1. **INTRODUCTION**

Some 3000 objects have been identified and cataloged since discovery of the first Kuiper Belt Object (KBO) in 1992 (Jewitt & Luu 1993). De-biased surveys estimate the Kuiper Belt region to contain ~$10^5$ objects having diameters larger than *d*~100 km (Petit et al. 2011). The physical properties of these objects provide important points for comparison and context between, and within, the various identified dynamical populations of the Kuiper Belt.

The New Horizons spacecraft, launched in January 2006, flew through the Pluto system, a large and highly complex member of the Kuiper Belt, in July 2015. On 1 January 2019 New Horizons will fly by a small KBO, (486958) 2014 MU$_{69}$ (Moore et al. 2018). During its first Kuiper Belt extended mission New Horizons will also observe ~25 typically-sized KBOs, Centaurs and dwarf planets at a variety of distances and phase angles (Stern et al. 2018). These observations are aimed at learning more about the shapes, satellite populations, surface scattering properties and other attributes of these KBOs in an effort to better understand the nature of KBOs and to place the flyby datasets of 2014 MU$_{69}$ in better context with respect to the larger KBO population (Rabinowitz 2007; Schaefer et al. 2009; Verbiscer 2018, Porter et al. 2016).

One of the outstanding questions about the Kuiper Belt revolves around the transition size where evolutionary collisional processes dominate over primordial surfaces. Studies of the size-distribution show that a break occurs at different absolute magnitude in each dynamical population of KBOs (Bernstein et al. 2004; Fuentes et al. 2010; Shankman et al. 2013; Adams et al. 2014; Fraser et al. 2014), which may be attributable to the same mechanism at those different diameters in the different regions. In the Cold Classical (CC) region of the belt — objects which have small eccentricities and inclinations and reside between the 3:2 and 2:1 mean motion resonance with Neptune — this break occurs near an absolute magnitude of $H_V$~7.2. All of the New Horizons extended mission targets lie beyond (i.e., smaller than) this break point. Work by Schlichting et al. (2013) and Trilling et al. (2006) support the idea that collisional grinding is responsible for population turnover, similar to what is observed in the main asteroid belt (Bottke et al. 2005). Recently, however, dynamical modeling by Shannon et al. (2016) and Singer et al. (2018) suggest that objects beyond the break are primordial in origin. Additional observations are necessary to characterize the physical properties of this population.

Colors and the fraction of binaries can speak to this question. Generally, the colors of KBOs are found to be diverse, from bluer than solar to redder than Mars (Benecchi et al. 2011; Fraser & Brown 2012; Hainaut et al. 2012; Fraser et al. 2015). Although mechanisms for surface processing in the Kuiper Belt are not well understood (Thébault 2008), it is generally thought that bluer objects have been more recently resurfaced than redder objects (Brunetto et al. 2007) and that reddening can happen quickly (Grundy 2009). The CCKBOs are statistically redder than other populations (Gulbis et al. 2006; Peixinho et al. 2008, 2015; Benecchi et al. 2011; Fraser et al. 2015) and have moderately high albedos (Brucker et al. 2009; Vilenius et al. 2014). They are thought to have formed near their current location, and not to have been substantially scattered, nor significantly collisionally processed during the giant planet migrations (Parker & Kavelaars 2010; Batygin et al. 2011; Volk & Malhotra 2011; Morbidelli et al. 2014).

Each dynamical region of the Kuiper Belt also appears to have its own binary fraction (Noll et al. 2008). The CCKBOs have a significantly higher apparent binary fraction, $29^{+7}_{-6}$% (Noll et al. 2008), than those of other dynamical populations, which are closer to ~15% (Noll et al. 2008). However, little work has been done with objects fainter than $H_V$~8 due to the difficulty of obtaining high signal to noise data for such small, faint objects. A high intrinsic fraction of binaries (≥30%) would support the idea that the size-distribution turnover is largely primordial while a low intrinsic fraction of binaries (≤15%) would indicate a significant role from processing by collisional grinding (Nesvorny et al. 2011). Nesvorny et al. also found that, with crude corrections for biasing, there is no evidence of a decrease in binary fraction at these smaller sizes. Due to geometric constraints our observations of 8 previously unobserved KBOs increases, but does not significantly further constrain, the underlying fraction of binaries for fainter/smaller KBOs.

In this paper we report optical color measurements obtained with the Hubble Space Telescope (HST) for 8 of the KBOs that New Horizons has, or will be observing. Three additional long-range targets were previously reported in Benecchi et al. 2015. These objects are not all inclusive of the distant KBOs to be observed by *New Horizons*, but are the objects which had been selected for distant observation at the time of the Cycle 23 proposal deadline and which did not have any previous HST color measurements. The overall F606W-F814W color for 1994 $JR_1$ from our dataset was included in Porter et al. 2016, however here we report the individual image photometry. We anticipate that these data will provide critical color information about these objects when they are observed by New Horizons' panchromatic LOng Range Reconnaissance Imager [LORRI; Cheng et al. 2008] to help place the spacecraft observations in proper context with other KBO datasets. We also investigate short term variability for these objects and binarity at the resolution of HST (~0.06 arcsec for these data).

## 2. OBSERVATIONS

The HST targets chosen for this program (GO 14092) included those selected for long-range observation by New Horizons prior to the HST proposal deadline for Cycle 23 and those not previously observed for color by HST. HST tracked these KBO targets at their rate of motion and obtained images using the Wide Field Camera 3 (WFC3) in the UVIS channel on HST (0.039621 arcsec/pixel, field of view 162 arcsec square) with the F606W filter, nominally centered at 600.1 nm with FWHM of 150.2 nm, comparable to a Johnson V filter, and the F814W filter, centered at 799.6 nm with FWHM 152.2 nm, comparable to Cousins I (Heyer et al. 2004). Observations for each target used a linear dither pattern and followed the sequence: F606W – F606W —F814W – F814W – F814W, with as many as 4 consecutive HST orbits following the same pattern for the faint objects. Switching between the two filters during each HST orbit enables checking for lightcurve effects which might negatively affect the accuracy of our color results. Exposures were 348 seconds in F606W and 373 seconds in F814W with the goal of obtaining S/N~5 for a single image, and color measurements good to ~0.1 to ~0.15 magnitudes with the combined images.

Table 1 summarizes the raw observations. Because these targets are in high-density star fields, HST observation times were pre-selected based on the best-known star catalogs to limit the probability of being close to, or overlapping, bright stars in the field of view. Fortuitously HST and New Horizons were able to observe 1994 $JR_1$ simultaneously. Figure 1 shows the location of these KBOs with respect to the background Classical Kuiper Belt in *a,e* and *i* space with 2014 $MU_{69}$ highlighted near the edge of the CC kernel. Figure 2 shows a sample HST image for each object in a single F606W integration.

INSERT TABLE 1 HERE
INSERT FIGURE 1 HERE
INSERT FIGURE 2 HERE

## 3. DETECTION AND PHOTOMETRIC ANALYSIS

All 8 KBOs were observed as close to opposition as possible to maximize their photometric signal to noise ratio. Each KBO target was detected in individual images, although the fainter targets were also studied by registering and stacking the images based on the expected plane of sky motion of the KBO. A handful of images (noted in Table 2 with dashes) were not used for photometry due to cosmic ray, signal dropouts or faint background sources that were undetected in the pre-field selection process We note that 2014 PN$_{70}$ likely has a significant lightcurve amplitude as its brightness changed consistently and by nearly a magnitude over the course of 4 consecutive HST orbits. 2014 OS$_{393}$ also has a significant lightcurve as it dropped below the signal threshold during the third visit in the F814W filter, but came back during the 4$^{th}$ visit (see section 4).

Photometry was carried out using an IDL PSF-Tiny Tim matching routine originally written by M. Buie, but adapted over the years by author Benecchi for program specific aspects. This methodology employs the amoeba (Press et al., 1992) routine, which performs multidimensional minimization of a function containing all our variables using the downhill simplex method, to optimize the fits. For each image we model both a single (x1,y1, flux1) and double PSF (x1,y1,flux1,x2,y2,flux2) and sky background. For 2014MU$_{69}$ we employed a slightly different methodology in which we force the pixel position of the object (x1,y1) from the now highly accurately determined orbit (Porter et al. 2018) and fit for only the flux and background values.

The calibrated Vega-magnitudes (Table 2), originally extracted from the image in the ST-magnitude system (STMAG), were derived from the observed counts matched to the data for an infinite aperture on the Tiny Tim models using the inverse sensitivity and photometric zero-point keyword values (PHOTFLAM and PHOTZPT) from the HST image headers (Rajan et al. 2011). This technique results in photometry extracted from noise-free images (corrected by the pixel area map) that removes the sky background noise contribution from the photometric uncertainty, a contribution that can be high for low signal-to-noise objects. To estimate uncertainty on the flux itself we re-run the data/model comparison with steps in flux to determine the value at which the $\chi^2$ residual of the image changes by 1-sigma. Conversion to Vega-magnitude (the reference system for most ground-based measurements) is 0.246 and 1.261 magnitudes brighter in F606W and F814W, respectively, than the ST-magnitude (http://www.stsci.edu/hst/wfc3/analysis/uvis_zpts/uvis1_infinite). F606W and F814W average magnitudes for each object were determined and the larger of either the combined uncertainties or the scatter in the values were used to set the magnitude uncertainties (the number of good images for each object is given in columns 2 and 4 in Table 3). Given our methodology we note that short duration lightcurves have the potential to increase the uncertainties on our photometry for multi-orbit objects. We designed our observation sequence to be able to recognize such variability and comment on these objects in section 4. Lastly, the colors and uncertainties were calculated with the standard uncertainties added in quadrature.

INSERT TABLE 2 HERE

## 4. COLOR, BINARITY, SIZE AND VARIABILITY FINDINGS

Table 3 lists the observed colors of the KBOs from this program. In Figure 3 we then compare this newly obtained data with 145 KBO colors that have been previously measured using the same HST filters (Benecchi et al. 2009; Fraser et al. 2012; Benecchi et al. 2015). The F606W-F814W colors (a proxy for V-I) plotted span a wide range from 0.5 < F606W-F814W < 1.3. All the KBOs in this program, with the exception of 1994JR$_1$ which is our only 3:2 mean motion resonant object, are at the red end of the range with F606W-F814W>0.8 (considering 1-sigma uncertainties). These red colors are consistent with their location in and near the CCKBO population, with 5 of them residing in the proposed Classical kernel, with 2014 MU$_{69}$ being the faintest and on the edge of this region. A weighted F606W-F814 mean for CC KBOs brighter and fainter than M$_{F606W}$=24.0 yields a difference of 0.097 magnitudes; the weighted means are 0.938±0.008 for M$_{F606W}$<24.0 (39 objects) and 1.035±0.022 for M$_{F606W}$≥24.0 (11 objects), respectively. However, the mean uncertainties on the measurements for these two samples are 0.09 and 0.11 magnitudes, respectively, so this result is an interesting trend, but not statistically significant. We also note that there is likely a bias against measuring blue objects in the M$_{F606W}$≥24.0 bin, because at this magnitude the S/N for blue KBOs decreases with increasing wavelength.

One way mitigate this is to investigate the full observational history of the objects with attempted measurements versus those with actual measurements. Fraser et al. 2012 state that of their full sample, 6 objects were not observed in either the optical or NIR images and photometry for 6 additional objects were excluded due to nearby stellar PSF contamination – these non-observations do not affect our statistics because the objects are not part of our sample. Eight objects were measured in the optical (F606W and F814W as used here) only, but not in the NIR due to pointing issues, again this is a non-issue for our statistics because we only use the optical colors. In Benecchi et al. 2009 all objects were measured in both filters. We exclude the colors from Benecchi et al. 2011 because the F555W filter was used instead of the F606W filter so the color values are not directly comparable, however, we note that there were no objects observed in F555W which were not observed in the F814W filter for this program; 7 of those objects were fainter than M$_{F555W}$=24.0.

INSERT TABLE 3 HERE
INSERT FIGURE 3 HERE

We also examined the individual KBO images, as well as the stacked images, to look for evidence of multiple systems by comparing the single and double PSF residual fit results. In all cases, the results from HST were negative. No companions are detected with separations of ~0.06 arcsec (~2000 km at 44 AU) and Δm ≤ 0.5. Closer or smaller undetected companions may yet be detectable with observations from New Horizons where the angular resolution exceeds that of HST (Parker et al. 2013b; Buie et al. 2018).

Looking only at the CCKBOs, Figure 4 shows the binary fraction versus the number of objects observed with HST as well as the semi-major axis to Hill radius characteristics of the binaries with well-characterized mutual orbits and physical characteristics. There are 13 binaries within the kernel region of the CC Kuiper belt with H$_v$ between 5.3-7.8 magnitudes, which yield a binary fraction of 29.5+13% for this particular sample. The apparent binary fraction drops off significantly beyond H$_V$ ~ 7.5; however, there are a number of observational selection effects that must be accounted for when assessing the trends in intrinsic binary fraction. For example, observing smaller (fainter) objects probes less and less of the available Hill radius of the objects because the spatial resolution remains the same for all objects at a given distance, given HST's fixed angular resolution.

INSERT FIGURE 4 HERE

If we assume that the distribution of the ratio of semi-major axis to Hill radius is constant for binaries of all sizes, then we can estimate the separation distribution (in Hill radii) at discovery for a binary of any size. Using the latest distribution of measured binary orbits (Grundy et al., submitted), we estimate that this distribution should follow an exponential distribution with an index of β~0.035. That is,

$$p(d_0/r_h) = \frac{1}{\beta} e^{\frac{1}{\beta} d_0/r_h}, \qquad (1)$$

where $d_0$ is the physical separation at discovery, and $r_h$ is the system Hill radius. Using this distribution and adopting typical albedos and densities for CC KBOs, we can estimate the fraction of binary systems that would be discoverable by HST at a fixed observational epoch for any H magnitude. We show this in Figure 4 where the red curve is the approximate odds that a binary system of a given H magnitude would be observable by HST in a single visit, given the exponential semi-major axis to Hill radius distribution and a uniform sampling of system orientations.

Additionally, if the system mass is distributed as a broken power law derived from the luminosity function of Fraser et al. (2014), then the apparent binary fraction will change with H magnitude even if the intrinsic binary fraction is constant. This is because binary systems appear brighter relative to single objects of the same mass, due to their greater surface area. Where the luminosity function is steep (and thus the mass function is steep), binaries will be strongly over-represented at any given H, resulting in a higher apparent binary fraction (defined in luminosity space) for a given intrinsic binary fraction (defined in mass space). However, where the slope is shallow, binaries will be more proportionally represented, resulting in more similar apparent and intrinsic binary fractions. At the region where the slope changes from steep to shallow, we expect the apparent binary fraction to drop from a high, over-represented value to a lower value closer to the intrinsic binary fraction.

Adopting a simple parameterization of both of these effects, we produce a model of expected apparent binary fraction vs. H curve that is consistent with the observed trends, assuming an underlying constant intrinsic binary fraction. In fact, we would only expect to have seen one binary fainter than H=8 given the sample of 14 objects, so seeing zero is therefore not statistically unreasonable. The blue curve represents the model apparent binary fraction assuming a constant 22% *intrinsic* binary fraction, the estimated HST detectability curve, the Fraser et al. (2014) luminosity function for CC KBOs, and an average brightening factor of 0.25 mags for binary systems over single objects of the same mass. The black curve represents the observed binary fraction for classical KBOs with a mean inclination of less than 5°. From this we conclude that given our sample does not indicate any evidence of evolution in binary fraction as a function of system H magnitude. Our lack of discovery of faint binaries does not necessarily mean that they don't exist, only that they are not as over-represented as brighter systems, and that they do not contain an excess of widely-separated binaries. We expect that there are a substantial number of binaries in more tightly bound orbits (the gray shaded region on the right panel of Figure 4) in this sample, beyond the grasp of HST but potentially within reach of the planned long-range New Horizons observations of several targets. A much larger statistically significant sample of systems fainter than the luminosity function break magnitude, ideally from a well characterized survey such as the Outer Solar System Origins Survey (OSSOS; Bannister et al. 2018) is required to further work on whether faint binaries remain consistent with a constant intrinsic binary fraction at all sizes. Our results continue to support the conclusion drawn by Nesvorny et al. (2011) that the rollover in the size distribution was not produced by disruptive collisions, but is instead a fossil remnant of the KBO formation process.

Using the formalism of Bowell et al. (1989) and an assumed geometric albedo of $\rho=0.12$ (Kovalenko et al. 2017), our long-range targets have diameters that range from 50-100 km. These objects are in the approximate size range where sizable amplitude lightcurves (>0.2 magnitudes) might be expected based on results from Trilling et al. (2006) and Benecchi & Sheppard (2013). Both 2014 $PN_{70}$ and 2014 $OS_{393}$ have significant photometric variations over the course of 4 consecutive HST orbits (Figure 5), however the signal-to-noise and time durations are not sufficient for further interpretation. Longer time baseline observations using the Subaru Hyper Suprime-Cam confirm large variations for these objects (Porter et al. 2017, 2018). The physical interpretation of large variations among small objects is preferably attributed to object elongation rather than albedo patches (Trilling et al. 2006) since mechanisms for large albedo variations are not highly plausible for objects of this size range. Some of these objects could be close or contact binaries (Benecchi & Sheppard 2013; Thirouin & Sheppard 2017), however, work by Thirouin & Sheppard (2018a,b) suggest a lower contact binary fraction among the CC population (10%) than has been found for other dynamical regions of the Kuiper Belt (eg. Resonant objects 40-50%). For one of our brighter objects, 2011 $HK_{103}$, we see systematic brightening, however our observation interval is a single HST so further analysis is limited. For most of the other objects we don't see systematic variations in magnitude during a single HST orbit or between consecutive orbits so at least at our sparsely sampled intervals we don't find these objects to have large amplitude lightcurves indicative of highly elongated shapes.

The lightcurve of 2014 $MU_{69}$ was investigated more extensively in a separate HST program (GO 14627) and will be presented in a companion paper, Benecchi et al. 2018b. The ~0.3 magnitude variation observed in the color dataset presented in this present paper is within the amplitude of variation found in the longer duration dataset when uncertainties are taken into consideration.

INSERT FIGURE 5 HERE

## 5. CONCLUSIONS

Using HST data we obtained, we have reported on the measured the color, binarity and photometric characteristics of eight (of the several dozen) faint KBOs that are being observed panchromatically from the New Horizons long-range (<0.3 AU) extended mission. We found that our 8 small KBOs are all red in color with 0.8 < F606W-F814W < 1.2, consistent with being members of the CC population and not interlopers.

The New Horizons observations will enable photometry at phase angles unobservable from the Earth, and lightcurves and satellite searches as well, all of which can be compared with the observations at low phase angle reported here. We expect to later be able to combine our observations and those from New Horizons to be able to better understand the entire CC population as a whole.

We do not identify any binary KBO targets. None of these KBOs appear to be binary seen from HST at the level of 0.06 arcsec separation (~2000 km) and $\Delta m<0.5$. However, New Horizons is flying through the kernel region of the CC Kuiper Belt where ~29% of objects are binary. All of our targets are below the size corresponding to $H_v>7.8$ and a few – 2014 PN$_{70}$ and 2014 OS$_{393}$ – were highly variable, so perhaps some of these objects are closer binaries than we can resolve, or are contact binaries. At <0.3 AU range New Horizons has ~5x the linear resolution and higher sensitivity than HST, and can detect companions unresolvable from HST. These results continue to support the conclusion drawn by Nesvorny et al. (2011) that the rollover in the CCKBO size distribution was not produced by disruptive collisions, but is instead a fossil remnant of the KBO formation process.

## ACKNOWLEDGMENTS

These observations used the NASA/ESA Hubble Space Telescope, and obtained at the Space Telescope Science Institute, which is operated by the Association of Universities for Research in Astronomy, Inc., under NASA contract NAS 5-26555. These observations are associated with programs 14092 and 13716. Support for both programs were provided by NASA through a grant from the Space Telescope Science Institute, which is operated by the Association of Universities for Research in Astronomy, Inc., under NASA contract NAS 5-26555. Some member of this team were also supported by NASA's New Horizons project. We thank Jean-Marc Petit and an anonymous reviewer for their helpful comments on this manuscript.

## REFERENCES

Adams, E. R., Gulbis, A. A. S., Elliot, J. L., Benecchi, S. D., Buie, M. W., Trilling, D. E., Wasserman, L. H. 2014. De-biased Populations of Kuiper Belt Objects from the Deep Ecliptic Survey. The Astronomical Journal, 148, Issue 3, article id. 55.

Batygin, K., Brown, M. E., & Fraser, W. C. 2011. Retention of a Primordial Cold Classical Kuiper Belt in an Instability-Driven Model of Solar System Formation. The Astrophysical Journal, 738, 13. doi:10.1088/0004-637X/738/1/13

Benecchi, S. D., Noll, K.~S., Grundy, W. M., Buie, M. W., Stephens, D. C., Levison, H. F. 2009. The correlated colors of transneptunian binaries. Icarus 200, 292.

Benecchi, S.D., Noll, K.S., Stephens, D.C., Grundy, W.M., Rawlins, J., 2011. Optical and infrared colors of transneptunian objects observed with HST. Icarus 213, 693–709. doi:10.1016/j.icarus.2011.03.005

Benecchi, S.D., Noll, K.S., Weaver, H.A., Spencer, J.R., Stern, S.A., Buie, M.W., Parker, A.H., 2015. New Horizons: Long-range Kuiper Belt targets observed by the Hubble Space Telescope. Icarus 246, 369–374. doi:10.1016/j.icarus.2014.04.014

Benecchi, S.D., Sheppard, S.S., 2013. Light Curves of 32 Large Transneptunian Objects. The Astronomical Journal 145, 124. doi:10.1088/0004-6256/145/5/124

Bannister, M. T., and 35 colleagues 2018. OSSOS. VII. 800+ Trans-Neptunian Objects —The Complete Data Release. The Astrophysical Journal Supplement Series 236, 18.

Benecchi, S.D. Porter, S., Spencer, J. R., Verbiscer, A. J., Noll, K. S., Stern, S. A., Buie, M. W., Zangari, A. M., and Parker, A. 2018. The HST Lightcurve of (486958) 2014 MU₆₉. Icarus, submitted.

Bernstein, G. M., Trilling, D. E., Allen, R. L., Brown, M. E., Holman, M., and Malhotra, R. 2004. The Size Distribution of Trans-Neptunian Bodies. The Astronomical Journal 128, 1364.

Bottke, W. F., Durda, D. D., Nesvorný, D. Jedicke, R., Morbidelli, A., Vokrouhlický, D., and Levison, Hal 2005. The fossilized size distribution of the main asteroid belt. Icarus 175, 111.

Bowell, E., Hapke, B., Domingue, D., Lumme, K., Peltoniemi, J., Harris, A.W., 1989. Application of photometric models to asteroids, in:. Presented at the IN: Asteroids II; Proceedings of the Conference, pp. 524–556.

Brucker, M.J., Grundy, W.M., Stansberry, J.A., Spencer, J.R., Sheppard, S.S., Chiang, E.I., Buie, M.W., 2009a. High albedos of low inclination Classical Kuiper belt objects. Icarus 201, 284–294. doi:10.1016/j.icarus.2008.12.040

Brunetto, R., de Le'on, J., Licandro, J. 2007. Testing space weathering models on A-type asteroid (1951) Lick. Astronomy and Astrophysics 472, 653.

Cheng, A.F., et al. 2008. Long-Range Reconnaissance Imager on New Horizons. Space Science Reviews 140, 189–215. doi:10.1007/s11214-007-9271-6

Fraser, W.C., Brown, M.E., 2012. The Hubble Wide Field Camera 3 Test of Surfaces in the Outer Solar System: The Compositional Classes of the Kuiper Belt. ApJ 749, 33. doi:10.1088/0004-637X/749/1/33

Fraser, W.C., Brown, M.E., Morbidelli, A., Parker, A., Batygin, K., 2014. The Absolute Magnitude Distribution of Kuiper Belt Objects. ApJ 782, 100. doi:10.1088/0004-637X/782/2/100

Fraser, W.C., Brown, M.E., and Glass, F. 2015. The Hubble Wide Field Camera 3 Test of Surfaces in the Outer Solar System: Spectral Variation on Kuiper Belt Objects. The Astrophysical Journal 804, 31.

Fuentes, C.I., Holman, M.J., Trilling, D.E., Protopapas, P., 2010. Trans-Neptunian Objects with Hubble Space Telescope ACS/WFC. ApJ 722, 1290–1302. doi:10.1088/0004-637X/722/2/1290

Grundy, W. M., Noll, K.S. , Roe H.G., Buie, M.W. Porter, S.B. Parker, A.H. Nesvorny, D., Levison, H.F., Benecchi, S.D., Stephens, D.C., and Trujillo C.A. 2018. Mutual Orbit Orientations of Transneptunian Binaries. Icarus, submitted.

Grundy, W. M. 2009.\Is the missing ultra-red material colorless ice?. Icarus 199, 560.

Gulbis, A.A.S., Elliot, J.L., Kane, J.F., 2006. The color of the Kuiper belt Core. Icarus 183, 168–178. doi:10.1016/j.icarus.2006.01.021

Hainaut, O. R., Boehnhardt, H., Protopapa, S. 2012. Colours of minor bodies in the outer solar system. II. A statistical analysis revisited. Astronomy and Astrophysics 546, A115.

Heyer, I., and 34 colleagues, 2004. WFPC2 Instrument Handbook: Version 9.0. STScI, Baltimore.

Jewitt, D., Luu, J., 1993. Discovery of the candidate Kuiper belt object 1992 QB1. Nature (ISSN 0028-0836) 362, 730–732. doi:10.1038/362730a0

Kovalenko, I. D., Doressoundiram, A., Lellouch, E., Vilenius, E., Muller, T., Stansberry, J. 2017. "TNOs are Cool": A survey of the trans-Neptunian region. XIII. Statistical analysis of multiple trans-Neptunian objects observed with Herschel Space Observatory. Astronomy and Astrophysics 608, A19.

Moore, J. M., McKinnon, W. B., Cruikshank, D. P., Gladstone, G. R., Spencer, J. R., Stern, S. A., et al. 2018. Great expectations: Plans and predictions for new horizons encounter with Kuiper Belt object 2014 MU69 ("Ultima Thule"). Geophysical Research Letters, 45. https://doi.org/10.1029/2018GL078996

Morbidelli, A., Gaspar, H.S., Nesvorný, D., 2014. Origin of the peculiar eccentricity distribution of the inner cold Kuiper belt. Icarus 232, 81–87. doi:10.1016/j.icarus.2013.12.023

Nesvorny, D., Vokrouhlicky, D., Bottke, W.F., Noll, K., Levison, H.F., 2011. Observed Binary Fraction Sets Limits on the Extent of Collisional Grinding in the Kuiper Belt. The Astronomical Journal 141, 159. doi:10.1088/0004-6256/141/5/159

Noll, K.S., Grundy, W.M., Chiang, E.I., Margot, J.L., Kern, S.D., 2008a. Binaries in the Kuiper Belt. The Solar System Beyond Neptune 345–363.

Noll, K.S., Grundy, W.M., Stephens, D.C., Levison, H.F., Kern, S.D., 2008b. Evidence for two populations of classical transneptunian objects: The strong inclination dependence of classical binaries. Icarus 194, 758–768. doi:10.1016/j.icarus.2007.10.022

Noll, K.S., Parker, A.H., Grundy, W.M., 2014. All Bright Cold Classical KBOs are Binary. American Astronomical Society 46.

Peixinho, N., Lacerda, P., & Jewitt, D. C. 2008. Color-Inclination Relation of the Classical Kuiper Belt Objects. The Astronomical Journal, 136, 1837. doi:10.1088/0004-6256/136/5/1837

Peixinho, N., Delsanti, A., and Doressoundiram, A. 2015. Reanalyzing the visible colors of Centaurs and KBOs: What is there and what we might be missing. Astronomy and Astrophysics 577, A35.

Petit, J.-M., Kavelaars, J.J., Gladman, B.J., Jones, R.L., Parker, J.W., van Laerhoven, C., Nicholson, P., Mars, G., Rousselot, P., Mousis, O., Marsden, B.G., Bieryla, A., Taylor, M., Ashby, M.L.N., Benavidez, P., Campo Bagatin, A., Bernabeu, G., 2011. The Canada-France Ecliptic Plane Survey—Full Data Release: The Orbital Structure of the Kuiper Belt. The Astronomical Journal 142, 131. doi:10.1088/0004-6256/142/4/131

Porter, S. B., and 8 colleagues. 2018. New Horizons Distant Observations of Cold Classical KBOs. AAS/Division for Planetary Sciences Meeting Abstracts 509.07.

Porter, S., and 7 colleagues. 2017. Constraints on the Shapes and Rotational States of the Distant New Horizons Kuiper Belt Targets. AGU Fall Meeting Abstracts 2017, P13F–P107.

Porter, S. B., Buie, M. W., Parker, A. H., Spencer, J. R., Benecchi, S., Tanga, P., Verbiscer, A., Kavelaars, J. J., Gwyn, S. D. J., Young, E. F., Weaver, H. A., Olkin, C. B., Parker, J. W., and Stern, S. A.. 2018. High-precision Orbit Fitting and Uncertainty Analysis of (486958) 2014 MU69. The Astronomical Journal 156, 20.

Porter, S. B., Spencer, J. R., Benecchi, S., Verbiscer, A. J., Zangari, A. M., Weaver, H. A., Lauer, T. R., Parker, A. H., Buie, M. W., Cheng, A. F., Young, L. A., Olkin, C. B., Ennico, K., Stern, S. A., the New Horizons Science Team. 2016. The First High-phase Observations of a KBO: New Horizons Imaging of (15810) 1994 JR1 from the Kuiper Belt. The Astrophysical Journal Letters, Volume 828, Issue 2, L15, 6 pp.

Press, W.H., Teukolsky, S.A., Vetterling, W.T., Flannery, B.P., 1992. Numerical Recipes in C, second ed. Cambridge University Press, New York.

Rabinowitz, D.L., Schaefer, B.E., Tourtellotte, S.W., 2007. The Diverse Solar Phase Curves of Distant Icy Bodies. I. Photometric Observations of 18 Trans-Neptunian Objects, 7 Centaurs, and Nereid. The Astronomical Journal 133, 26. doi:10.1086/508931

Rajan, A. et al. 2011. "WFC3 Data Handbook", Version 2.1, (Baltimore: STScI). http://www.stsci.edu/hst/wfc3/documents/handbooks/currentDHB/wfc3_dhb.pdf

Schaefer, B.E., Rabinowitz, D.L., Tourtellotte, S.W., 2009. The Diverse Solar Phase Curves of Distant Icy Bodies II. The Cause of the Opposition Surges and Their Correlations. The Astronomical Journal 137, 129. doi:10.1088/0004-6256/137/1/129

Schlichting, H. E., Fuentes, C. I., & Trilling, D. E. 2013. Initial Planetesimal Sizes and the Size Distribution of Small Kuiper Belt Objects. The Astronomical Journal, 146(2), 36. doi:10.1088/0004-6256/146/2/36

Shankman, C., Gladman, B.J., Kaib, N., Kavelaars, J.J., Petit, J.-M., 2013. A Possible Divot in the Size Distribution of the Kuiper Belt's Scattering Objects. ApJ 764, L2. doi:10.1088/2041-8205/764/1/L2

Shannon, A., Wu, Y., Lithwick, Y., 2016. Forming the Cold Classical Kuiper Belt in a Light Disk. ApJ 818, 175. doi:10.3847/0004-637X/818/2/175

Stern et al. 2018, Space Science Revs, in press.

Thébault, P., 2004. A Numerical Check of the Collisional Resurfacing Scenario, in: The First Decadal Review of the Edgeworth-Kuiper Belt. Springer Netherlands, Dordrecht, pp. 233–241. doi:10.1007/978-94-017-3321-2_21

Thirouin, A., Sheppard, S. S. 2018a. Lightcurves of the Dynamically Cold Classical Trans-Neptunian Objects. AAS/Division for Planetary Sciences Meeting Abstracts 302.05.

Thirouin, A., Sheppard, S. S. 2018b. The Plutino Population: An Abundance of Contact Binaries. The Astronomical Journal 155, 248.

Thirouin, A., Sheppard, S. S. 2017. A Possible Dynamically Cold Classical Contact Binary: (126719) 2002 CC249. The Astronomical Journal 154, 241.

Trilling, D.E., Bernstein, G.M., 2006. Light Curves of 20-100 km Kuiper Belt Objects Using the Hubble Space Telescope. The Astronomical Journal 131, 1149–1162. doi:10.1086/499228

Verbiscer, A.J., and 13 colleagues. 2018. Phase Curves of Nix and Hydra from the New Horizons Imaging Cameras. The Astrophysical Journal 852, L35.

Volk, K., & Malhotra, R. 2011. Inclination Mixing in the Classical Kuiper Belt. The Astrophysical Journal, 736, 11. doi:10.1088/0004-637X/736/1/11

# TABLES

TABLE 1. OBSERVATIONS

| Target | # HST Orbits | $H_v$ | Data Collection (UT) | Class[b] | R (AU)[d] | Δ (AU)[d] | α(°)[d] | a (AU) | e | $i_{mean}$ |
|---|---|---|---|---|---|---|---|---|---|---|
| 1994JR1[a] | 1 | 7.70 | 2015-11-02 05:24-06:06 | 3:2 | 35.46 | 35.91 | 1.43 | 39.614 | 0.123 | 3.428 |
| 2011HK103 | 1 | 8.42 | 2016-06-18 14:22-15:03 | SN | 42.91 | 41.93 | 0.38 | 53.579 | 0.313 | 5.256 |
| 2011HF103 | 1 | 8.45 | 2016-06-19 14:13-14:54 | CC | 43.04 | 42.06 | 0.36 | 43.265 | 0.052 | 3.084 |
| 2011JW31 | 2 | 9.28 | 2016-07-06 13:13-15:51 | CC | 42.67 | 41.65 | 0.05 | 45.861 | 0.098 | 3.158 |
| 2012HE85 | 2 | 9.04 | 2016-07-05 13:22-16:00 | 9:5 | 40.22 | 39.21 | 0.06 | 44.801 | 0.104 | 4.662 |
| 2014OS393 | 4 | 10.07 | 2016-06-22 12:10-17:38 | CC | 43.34 | 42.34 | 0.29 | 44.427 | 0.063 | 2.911 |
| 2014PN70 | 4 | 10.16 | 2016-07-09 14:21-20:09 | CC | 44.04 | 43.03 | 0.11 | 44.454 | 0.055 | 3.272 |
| 2014MU69[c] | 4 | 11.10 | 2016-07-03 13:48-19:16 | CC | 43.33 | 42.31 | 0.06 | 44.493 | 0.04 | 2.521 |

[a] simultaneous with New Horizons observations. [b] dynamical classification based on the work of Elliot et al. 2005; SN = Scattered Near (nonresonant, non–planet-crossing objects characterized by time-averaged Tisserand parameters less than 3, relative to Neptune), CC = Cold Classical (KBOs with mean Tisserand parameters greater than 3, time-averaged eccentricities less than 0.2 and an inclination <5°). [c] New Horizons primary fly-by target. [d] R=Heliocentric distance, Δ = Geocentric distance and α=phase angle at time of HST observation.

TABLE 2. INDIVIDUAL IMAGE PHOTOMETRY

| Object | Image | JD$_{Midtime}$ | Filter | Vega-Magnitude |
|---|---|---|---|---|
| 1994JR1 | icze01egq | 2457328.72766 | F606W | 22.73±0.03 |
| 1994JR1 | icze01ehq | 2457328.73355 | F606W | 22.65±0.03 |
| 1994JR1 | icze01ejq | 2457328.73983 | F814W | 21.97±0.03 |
| 1994JR1 | icze01emq | 2457328.74601 | F814W | 21.94±0.03 |
| 1994JR1 | icze01eqq | 2457328.75219 | F814W | 22.01±0.03 |
| 2011HK103 | icze02hsq | 2457558.10083 | F606W | 23.78±0.03 |
| 2011HK103 | icze02htq | 2457558.10672 | F606W | 23.57±0.02 |
| 2011HK103 | icze02hvq | 2457558.11303 | F814W | 22.86±0.03 |
| 2011HK103 | icze02hxq | 2457558.11921 | F814W | 22.77±0.03 |
| 2011HK103 | icze02hzq | 2457558.12539 | F814W | 22.68±0.03 |
| 2011HF103 | icze03n0q | 2457559.09446 | F606W | 23.98±0.03 |
| 2011HF103 | icze03n1q | 2457559.10035 | F606W | 24.04±0.03 |
| 2011HF103 | icze03n3q | 2457559.10665 | F814W | 23.03±0.04 |
| 2011HF103 | icze03n5q | 2457559.11283 | F814W | 23.04±0.04 |
| 2011HF103 | icze03n7q | 2457559.11901 | F814W | 22.98±0.03 |
| 2011JW31 | icze04usq | 2457576.05885 | F606W | 23.98±0.03 |
| 2011JW31 | icze04utq | 2457576.06475 | F606W | 23.92±0.03 |
| 2011JW31 | icze04uvq | 2457576.07105 | F814W | 22.86±0.03 |
| 2011JW31 | icze04uxq | 2457576.07723 | F814W | 22.85±0.03 |
| 2011JW31 | icze04uzq | 2457576.08341 | F814W | 22.92±0.03 |
| 2011JW31 | icze05v1q | 2457576.12512 | F606W | 24.08±0.03 |
| 2011JW31 | icze05v2q | 2457576.13101 | F606W | 24.14±0.03 |
| 2011JW31 | icze05v4q | 2457576.13731 | F814W | 22.97±0.03 |
| 2011JW31 | icze05v6q | 2457576.14349 | F814W | 23.08±0.04 |
| 2011JW31 | icze05v8q | 2457576.14967 | F814W | 23.10±0.04 |
| 2012HE85 | icze06o6q | 2457575.06523 | F606W | 24.51±0.04 |
| 2012HE85 | icze06o7q | 2457575.07112 | F606W | 24.54±0.04 |
| 2012HE85 | icze06o9q | 2457575.07742 | F814W | 23.50±0.05 |
| 2012HE85 | icze06obq | 2457575.08361 | F814W | 23.34±0.04 |
| 2012HE85 | icze06odq | 2457575.08979 | F814W | 23.55±0.05 |
| 2012HE85 | icze07ofq | 2457575.13147 | F606W | 24.46±0.04 |
| 2012HE85 | icze07ogq | 2457575.13736 | F606W | 24.48±0.04 |
| 2012HE85 | icze07oiq | 2457575.14366 | F814W | 23.43±0.04 |
| 2012HE85 | icze07okq | 2457575.14984 | F814W | 23.39±0.04 |
| 2012HE85 | icze07omq | 2457575.15602 | F814W | 23.38±0.04 |

| **Object** | **Image** | **JD_Midtime** | **Filter** | **Vega-Magnitude** |
|---|---|---|---|---|
| 2014OS393 | icze08tlq | 2457562.00965 | F606W | — |
| 2014OS393 | icze08tmq | 2457562.01554 | F606W | 25.56±0.07 |
| 2014OS393 | icze08toq | 2457562.02185 | F814W | 24.69±0.10 |
| 2014OS393 | icze08tqq | 2457562.02803 | F814W | 24.43±0.08 |
| 2014OS393 | icze08tsq | 2457562.03421 | F814W | 24.40±0.08 |
| 2014OS393 | icze09tuq | 2457562.07593 | F606W | 25.57±0.07 |
| 2014OS393 | icze09tvq | 2457562.08182 | F606W | 25.53±0.07 |
| 2014OS393 | icze09txq | 2457562.08812 | F814W | 24.31±0.08 |
| 2014OS393 | icze09tzq | 2457562.09430 | F814W | 24.29±0.08 |
| 2014OS393 | icze09u1q | 2457562.10048 | F814W | 24.81±0.11 |
| 2014OS393 | icze10u3q | 2457562.14219 | F606W | 25.44±0.07 |
| 2014OS393 | icze10u4q | 2457562.14808 | F606W | 25.33±0.06 |
| 2014OS393 | icze10u6q | 2457562.15438 | F814W | Signal drop-out |
| 2014OS393 | icze10u8q | 2457562.16056 | F814W | Signal drop-out |
| 2014OS393 | icze10uaq | 2457562.16674 | F814W | Signal drop-out |
| 2014OS393 | icze11ucq | 2457562.20844 | F606W | 25.64±0.08 |
| 2014OS393 | icze11udq | 2457562.21433 | F606W | 25.57±0.07 |
| 2014OS393 | icze11ufq | 2457562.22063 | F814W | 25.04±0.13 |
| 2014OS393 | icze11uhq | 2457562.22681 | F814W | — |
| 2014OS393 | icze11ujq | 2457562.23299 | F814W | 24.64±0.09 |
| 2014PN70 | icze12kiq | 2457579.10575 | F606W | 26.46±0.14 |
| 2014PN70 | icze12kjq | 2457579.11164 | F606W | 25.58±0.07 |
| 2014PN70 | icze12klq | 2457579.11795 | F814W | 25.24±0.15 |
| 2014PN70 | icze12knq | 2457579.12413 | F814W | 24.78±0.10 |
| 2014PN70 | icze12kpq | 2457579.13031 | F814W | 25.13±0.13 |
| 2014PN70 | icze13krq | 2457579.17200 | F606W | 25.97±0.10 |
| 2014PN70 | icze13ksq | 2457579.17789 | F606W | 25.98±0.10 |
| 2014PN70 | icze13kuq | 2457579.18420 | F814W | 25.25±0.15 |
| 2014PN70 | icze13kwq | 2457579.19038 | F814W | — |
| 2014PN70 | icze13kyq | 2457579.19656 | F814W | 24.55±0.09 |
| 2014PN70 | icze14l0q | 2457579.23823 | F606W | 25.34±0.06 |
| 2014PN70 | icze14l1q | 2457579.24412 | F606W | 25.72±0.08 |
| 2014PN70 | icze14l3q | 2457579.25042 | F814W | 24.65±0.10 |
| 2014PN70 | icze14l5q | 2457579.25660 | F814W | 24.29±0.08 |
| 2014PN70 | icze14l7q | 2457579.26278 | F814W | 24.42±0.08 |
| 2014PN70 | icze15l9q | 2457579.30447 | F606W | 25.72±0.08 |
| 2014PN70 | icze15laq | 2457579.31036 | F606W | 25.58±0.07 |
| 2014PN70 | icze15lcq | 2457579.31666 | F814W | 24.75±0.10 |

| Object | Image | JD$_{Midtime}$ | Filter | Vega-Magnitude |
|---|---|---|---|---|
| 2014PN70 | icze15leq | 2457579.32284 | F814W | 24.69±0.10 |
| 2014PN70 | icze15lgq | 2457579.32902 | F814W | — |
| 2014MU69 | icze16jjq | 2457573.07750 | F606W | 26.04±0.10 |
| 2014MU69 | icze16jkq | 2457573.08339 | F606W | 25.91±0.09 |
| 2014MU69 | icze16jmq | 2457573.08969 | F814W | — |
| 2014MU69 | icze16joq | 2457573.09587 | F814W | 25.43±0.17 |
| 2014MU69 | icze16jqq | 2457573.10205 | F814W | 25.48±0.17 |
| 2014MU69 | icze17jsq | 2457573.14376 | F606W | 26.18±0.11 |
| 2014MU69 | icze17jtq | 2457573.14965 | F606W | 26.19±0.11 |
| 2014MU69 | icze17jvq | 2457573.15595 | F814W | 25.13±0.13 |
| 2014MU69 | icze17jxq | 2457573.16214 | F814W | 24.98±0.12 |
| 2014MU69 | icze17jzq | 2457573.16832 | F814W | — |
| 2014MU69 | icze18k1q | 2457573.21002 | F606W | 26.39±0.13 |
| 2014MU69 | icze18k2q | 2457573.21591 | F606W | 26.70±0.16 |
| 2014MU69 | icze18k4q | 2457573.22222 | F814W | 25.67±0.20 |
| 2014MU69 | icze18k6q | 2457573.22840 | F814W | 24.93±0.12 |
| 2014MU69 | icze18k8q | 2457573.23458 | F814W | 25.13±0.13 |
| 2014MU69 | icze19kaq | 2457573.27627 | F606W | 26.54±0.15 |
| 2014MU69 | icze19kbq | 2457573.28216 | F606W | 26.31±0.12 |
| 2014MU69 | icze19kdq | 2457573.28847 | F814W | 25.35±0.16 |
| 2014MU69 | icze19kfq | 2457573.29465 | F814W | 25.33±0.16 |
| 2014MU69 | icze19khq | 2457573.30083 | F814W | 25.11±0.13 |

TABLE 3. COLOR RESULTS

| Object | JDMidtime | #imgs | F606W | #imgs | F814W | F606W-F814 | V | I | V-I |
|---|---|---|---|---|---|---|---|---|---|
| 1994JR1[a] | 2457328.73992 | 2 | 22.69±0.04 | 3 | 21.98±0.02 | 0.72±0.03 | 23.00 | 21.96 | 1.04 |
| 2011JY31[b] | 2456189.24978 | 2 | 24.65±0.03 | 3 | 23.69±0.16 | 0.96±0.16 | 24.96 | 23.67 | 1.28 |
| 2011HZ102[b] | 2456188.31616 | 2 | 25.70±0.02 | 3 | 24.66±0.07 | 1.04±0.07 | 26.01 | 24.64 | 1.36 |
| 2013LU35[b] | 2456563.37731 | 2 | 26.34±0.09 | 3 | 25.23±0.02 | 1.11±0.09 | 26.65 | 25.21 | 1.43 |
| 2011HK103 | 2457558.11149 | 2 | 23.67±0.12 | 3 | 22.77±0.06 | 0.90±0.13 | 23.98 | 22.75 | 1.23 |
| 2011HF103 | 2457559.10512 | 2 | 24.01±0.03 | 3 | 23.02±0.02 | 1.00±0.04 | 24.32 | 23.00 | 1.32 |
| 2011JW31 | 2457576.10426 | 4 | 24.03±0.06 | 6 | 22.96±0.05 | 1.07±0.08 | 24.34 | 22.95 | 1.39 |
| 2012HE85 | 2457575.11063 | 4 | 24.50±0.05 | 6 | 23.43±0.04 | 1.07±0.04 | 24.80 | 23.42 | 1.39 |
| 2014OS393 | 2457562.12427 | 7 | 25.52±0.04 | 8 | 24.58±0.10 | 0.94±0.11 | 25.82 | 24.64 | 1.18 |
| 2014PN70 | 2457579.21430 | 8 | 25.79±0.13 | 10 | 24.78±0.11 | 1.02±0.17 | 26.10 | 24.76 | 1.34 |
| 2014MU69 | 2457573.18916 | 8 | 26.28±0.10 | 10 | 25.25±0.10 | 1.03±0.14 | 26.59 | 25.24 | 1.35 |

[a] From Porter et al. 2016. [b] From Benecchi et al. 2015

# FIGURES

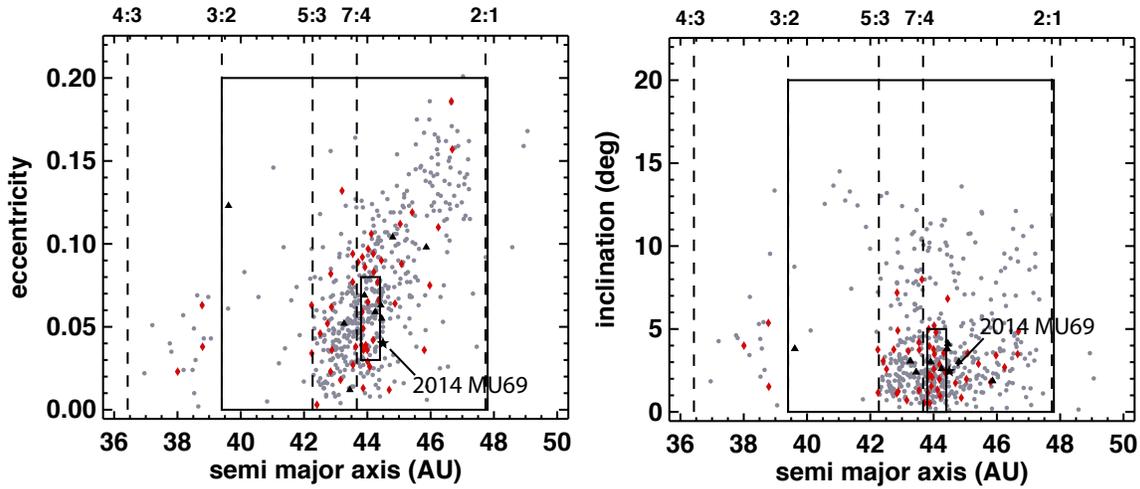

**Figure 1.** The location of the objects from this work and Benecchi et al. (2015) are highlighted in as black triangles with respect to the background Classical Kuiper Belt in *a,e* and *i* space. For clarity, no Resonant or Scattered objects are plotted, although the Resonant locations are show as vertical dashed lines. For reference binaries are highlighted in red and the formal locations of the Classical and kernel regions are highlighted in black boxes using the definitions from Petit et al. 2011). 2014 MU₆₉ is plotted as a black star and resides on the outside edge of the kernel.

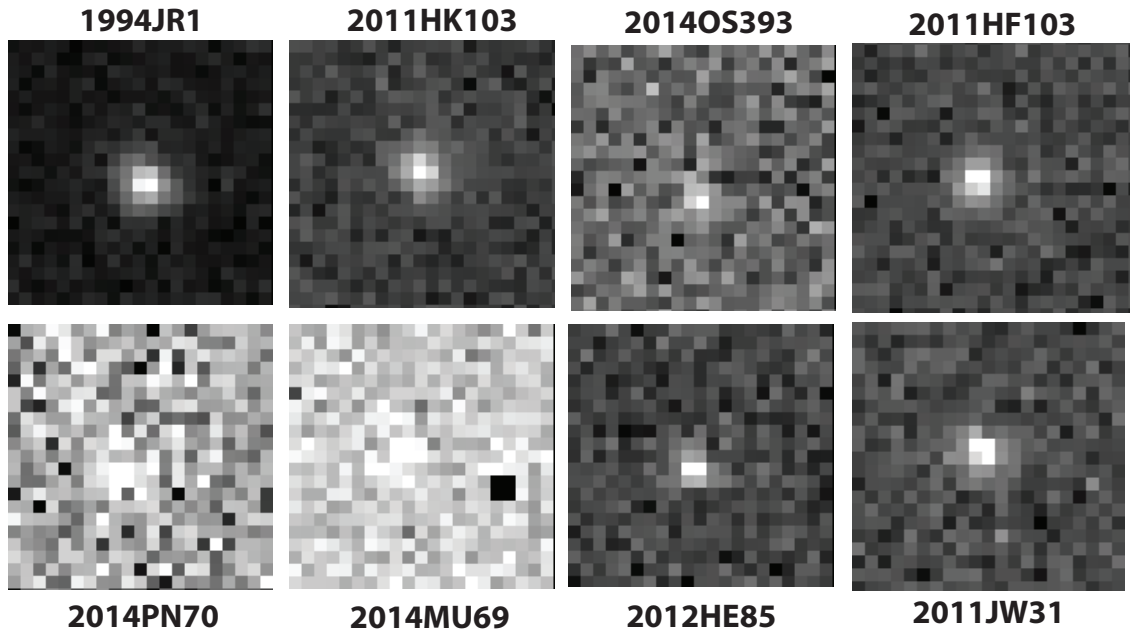

**Figure 2.** Sample HST images (pipeline calibrated, but otherwise raw format) in the F606W filter (comparable to the Johnson-V filter) for each long-range observation KBO presented in this paper.

Magnitudes in the F606W filter range from the brightest, 1994 JR₁ at m$_v$~23.0 to the faintest, 2014 MU₆₉ at m$_v$~26.6.

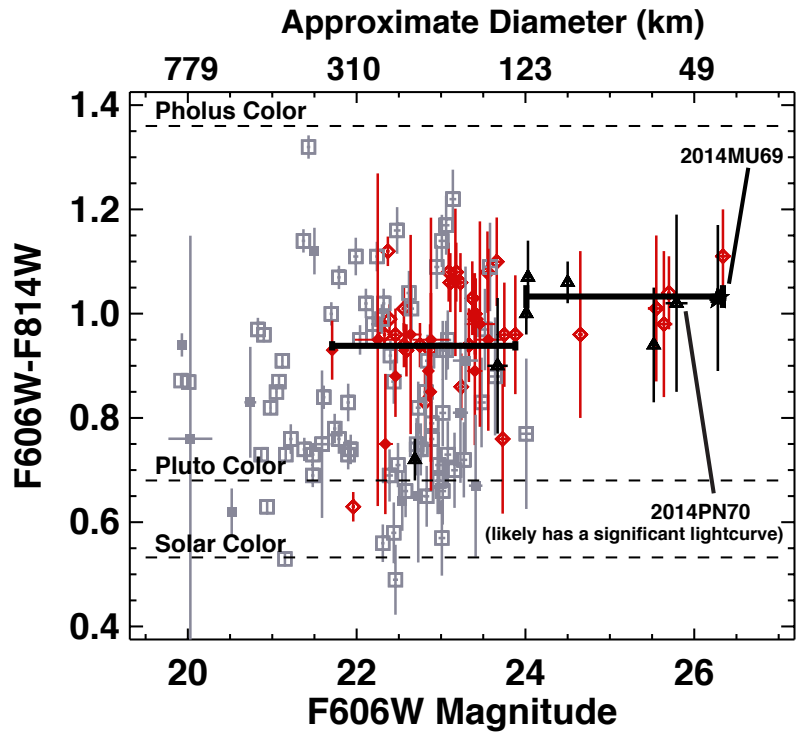

**Figure 3.** Color results for all long-range New Horizons targets with HST colors compared with the colors of all other KBOs that have been measured in the same HST filters (from Benecchi et al. 2009, Fuentes et al. 2010; Fraser et al. 2012 and Benecchi et al. 2015) with respect to their direct F606W magnitudes. Plotted along the top is an estimate of the effective diameters for these objects following the formulation of Bowell et al. (1989) d = 10((6.259−0.4·H$_{F606W}$ −log ρ)/2) km and assuming an albedo, ρ; we assume ρ = 0.12. The F606W-F814W color is similar to ground based Johnson V-Cousins I. Objects plotted as red diamonds are dynamically part of the CC population, objects in all other classes (Scattered, Resonant, hot Classical and Centaur combined) are plotted as gray squares. Binaries are plotted as filled symbols in their respective classifications. The color of the Sun in the HST filter set is plotted as a dashed line. The New Horizons objects observed in this program are plotted as black triangles and 2014 MU₆₉ is plotted as a solid black star. These objects are fainter and comparatively redder than many KBOs. A weighted F606W-F814 mean for CC KBOs brighter and fainter than M$_{F606W}$=24.0 yields a difference of 0.097 magnitudes; the two values are plotted as solid black lines with an average of 0.938±0.008 for M$_{F606W}$<24.0 (39 objects) and 1.035±0.022 for M$_{F606W}$≥24.0 (11 objects). We note that there is likely a bias against measuring blue objects in the M$_{F606W}$≥24.0 bin, because at this magnitude the S/N for blue KBOs decreases with increasing wavelength. Likewise, although the S/N is less for the fainter objects, 2014PN₇₀ appears to have strong lightcurve effects as it changes significantly and somewhat systematically during the 4 consecutive HST orbits (see Figure 5).

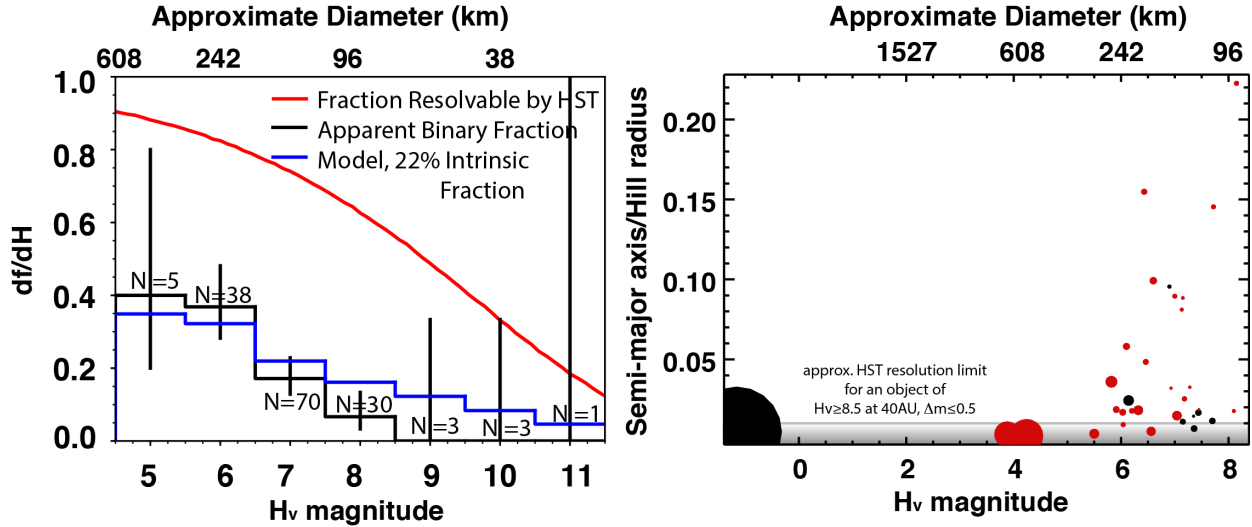

**Figure 4.** (Left) Binary fraction for CCKBOs (Noll et al. 2008, Nesvorny et al. 2011, Noll et al. 2014) vs. the number of CCKBOs observed with HST. The red curve is the odds that a binary system of a given H magnitude would be observable by HST in a single visit, given the observed semi-major axis to Hill radius distribution and a uniform sampling of system orientations. Bars represent 1 sigma uncertainties. The blue curve is the model apparent binary fraction assuming a constant 22% intrinsic binary fraction and selection effects described in the text. The black curve is the observed binary fraction for classical KBOs with a mean inclination of less than 5°. While the fraction drops off with $H_v$ magnitude. The number of known CC KBOs in the faintest bins is small because obtaining good heliocentric orbits on faint objects from ground-based observatories is extremely difficult. (Right) Binary $H_v$ magnitude vs. Semi-major axis/Hill radius ratio. Red objects are CCKBOs and the size of the circle is proportional to the size of the object. The gray shaded region gives the approximate HST resolution limit for an object of $H_v \geq 8.5$ at 40 AU with a component delta magnitude of ≤0.5 magnitudes. Because of their smaller Hill radii and the strong bias to small semi-major axes (less than 1% of Hill radius) CCKBO binaries are less likely to be resolvable, although many of these may be elongated or contact binary in nature (Buie et al. 2018). Thirteen binaries are within the kernel region of the Classical Kuiper belt with all of these objects having $H_v$ between 5.3-7.8 magnitudes yielding a (biased) binary fraction of 29.5±13%.

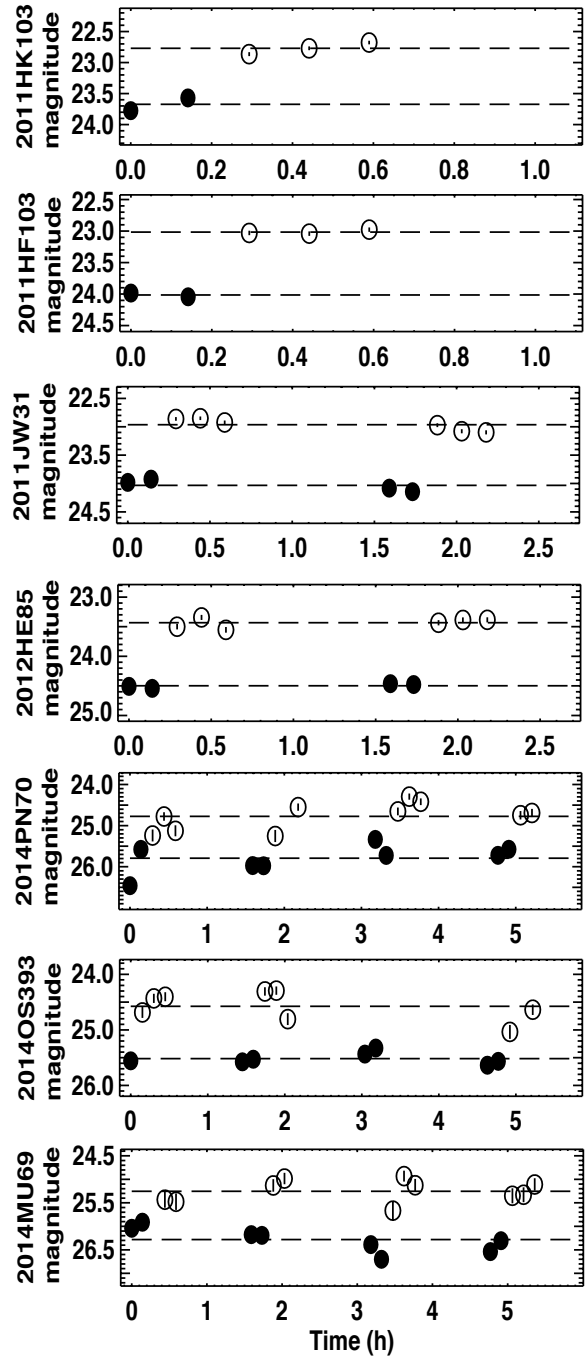

**Figure 5.** Variation of each object we observed vs. time, ordered by reflectance. The first two objects were observed within one HST orbit (~40 minutes) the second two over two consecutive HST orbits (spanning ~2 hours) and the last three over 4 consecutive HST orbits (spanning ~5 hours). The open and filled circles are measurements in the F814W and F606W filters, respectively. Error bars are shown as vertical lines inside the circles, if they are not visible they are smaller than the point size. We notice larger variability in the fainter objects independent of the observations being seen at lower signal-to-noise ratios, although the data are not of sufficient quality to be further interpreted.